\documentclass[12pt]{iopart}
\usepackage{iopams}  
\begin{document}

\title{A note on confined diffusion}

\author{Thomas Bickel}

\address{Group of Statistical Physics, CPMOH \& Universit\'e Bordeaux 1 (UMR 5798) \\
351 cours de la Lib\'eration, 33405 Talence, France }
\ead{th.bickel@cpmoh.u-bordeaux1.fr}

\begin{abstract}

The random motion of a Brownian particle confined in some finite domain
is considered. Quite generally, the relevant statistical properties
involve infinite series,
whose coefficients are related to the eigenvalues 
of the diffusion operator.
Unfortunately, the latter depend  on  space dimensionality 
and on the particular shape of the domain, and
an analytical expression 
is in most circumstances not available. 
In this article, it is shown that the series may in some circumstances
sum up exactly. 
Explicit calculations are performed for 2D diffusion
restricted to a circular domain and 3D diffusion inside a sphere.
In both cases, the short-time behaviour of 
the mean square displacement is obtained.\\


\end{abstract}

\maketitle

\section{Introduction}
\label{intro}

Although the theory of Brownian motion has been one of the cornerstones of
modern physics for more than one century~\cite{chandra}, it still raises numerous
fundamental questions~\cite{hanggi}. Applications are found in various disciplines,  especially
in biophysics where technical progress
allows nowadays for the detection of {\it individual} nanoparticles in living systems~\cite{cognet}.
In many instances, the particles' motion  is strictly restricted to bounded 
domains.  One can quote  corralled motion of receptors on cell membrane~\cite{saxton,
daumas,murase},
protein diffusion inside the cell nucleus~\cite{kues}, or transporter diffusion through 
nuclear pores~\cite{chatenay,bickelbruinsma}. In a typical experiment,  trajectories of tracer particles 
are recorded and analysed in terms of position correlations 
or mean square displacement.
From those datas, information on the size of the diffusing objects or on
their interactions with the surrounding medium can then be extracted~\cite{urbach}.

In this article, we consider the motion of a Brownian particle confined in a domain $\mathcal{D}$ of  given size
 and shape. The statistical properties of the particle are described by the Green's function
$G\left( \mathbf{r}, \mathbf{r}', t \right)$,
which represents the probability
density of finding the particle at point $\mathbf{r}$ and  time $t$,
together with the intial condition 
$G\left( \mathbf{r}, \mathbf{r}', 0 \right)=\delta \left( \mathbf{r} - \mathbf{r}' \right) $.
The Green's function satisfies the Fokker-Planck equation
\begin{equation}
\frac{\partial G}{\partial t} - D \nabla^2 G = \delta (\mathbf{r}-\mathbf{r}') \delta (t) 
\label{fpeq}
\end{equation}
with $D$ the diffusion coefficient. 
Confinement is enforced through the reflecting boundary
conditions on the frontier $\partial \mathcal{D}$ of the domain~\cite{chandra}. 
It should be noticed that, because of the finite size of the system, translational invariance is broken
and the Green's function depends on $\mathbf{r}$ and $\mathbf{r}'$ separately.
Once the propagator is known,
all the statistical properties can, in principle, be evaluated. For instance, 
the mean square displacement (MSD) is obtained from
\begin{eqnarray}
\left\langle \delta \mathbf{r}^2 (t) \right\rangle & = &  
\left\langle \left( \mathbf{r} (t) -\mathbf{r} (0) \right)^2 \right\rangle  \nonumber \\
& = &   \int \rmd^d \mathbf{r} \int  \rmd^d \mathbf{r}'  
\  G  \left(\mathbf{r}, \mathbf{r}', t  \right) P_0(\mathbf{r}')
 \left( \mathbf{r} -\mathbf{r}' \right)^2
\label{msd}
\end{eqnarray}
where $d$ is the dimension of the embedding space. The initial probability distribution is assumed 
to be uniform, $P_0(\mathbf{r}')= V_d^{-1}$,
with $V_d$ the  volume of the domain.

Of particular interest is the short-time behaviour of the MSD. Let us define 
$\tau$ the relaxation time of the system.
Simple dimensional analysis shows that $\tau$ scales as the square of 
the typical size of the domain divided by the diffusion coefficient.
It appears intuitively clear that for $t \ll \tau$, the particle 
has not enough time to feel the influence of the boundary~\cite{kac}.
The expected behaviour for the MSD is then $\langle \delta \mathbf{r}^2 (t) \rangle  \sim  2dDt  $,
but explicit derivation of this limit is however not always straightforward.
Indeed, the solution of the diffusion equation is usually written as an infinite series,
whose coefficients are related to the eigenvalues of the Laplacian.
The latter actually depend on  space dimensionality and on the particular shape
of the domain. Unfortunately, 
an analytical expression for the eigenvalues
is in most circumstances not available ---  
with the notable exception of the problem in 1D. 
In this article, we focus on 2D diffusion in a circular domain and 3D diffusion inside a sphere.
In both cases, the MSD is written as a sum of reciprocal powers of zeros
of derivatives of Bessel functions. The goal of this note  
is to derive explicitly the short-time limit of the MSD. 
Though this point might sound rather academic
--- since the result is known {\it a priori} ---,
it certainly deserves some comments.
To the best of the author's knowledge,
this questions has  never been raised before.
The derivation outlined in this article is based on the expansion of  entire functions as
an infinite product~\cite{arfkenbook}. This approach, originally developed by
Euler and Rayleigh~\cite{watsonbook}, has been extended recently by Muldoon and collaborators
to derive convolution formulas for Rayleigh functions~\cite{muldoon1,muldoon2}.
Adapting this method to the relevant Bessel  (or related) functions allows us to  
recover the short-time limit of the MSD.
The remaining  of  the paper is organized as follows. We first recall
in Section~\ref{secline} some classical results concerning 1D diffusion.
Then we present the calculation of the MSD on a disk in section~\ref{seccircle} and
inside a sphere in section~\ref{secsphere}. Some concluding remarks are finally drawn
in the last section. In order to be complete, we give in the Appendix the 
corresponding Laplace transforms of the MSD.

%
%

\section{Diffusion on a line}
\label{secline}

To illustrate our point, we first consider 1D diffusion restricted to a segment of length $L=2a$. 
The Green's function $G\left(x, x' , t \right)$ satisfies the diffusion equation~(\ref{fpeq})
together with the boundary conditions 
\begin{equation}
\frac{\partial G}{ \partial x} (0,x',t) = \frac{\partial G}{ \partial x} (L,x',t)=0 \ .
\end{equation}
Defining the relaxation time $\tau = a^2 /D$, the solution is written as a 
series~\cite{carslawbook}
\begin{equation}
 G (x,x',t)= \frac{1}{L}+\frac{2}{L} \sum_{n=1}^{\infty} 
\exp\left[ - \left( \frac{n \pi }{2} \right)^2 \frac{t}{\tau} \right] 
\cos  \left( \frac{n \pi x}{L} \right) \cos  \left( \frac{n \pi x'}{L} \right) \ .
\label{soldiff1D}
\end{equation}
Accordingly, the  MSD  $\langle \delta x^2 (t) \rangle = \langle (x(t) - x(0))^2 \rangle$   reads
\begin{eqnarray}
\langle \delta x^2 (t) \rangle & = &  \frac{1}{L} \int_0^L \rmd x \int_0^L \rmd x'  G (x,x',t) (x-x')^2  \nonumber \\
& = &  \frac{L^2}{6} \left( 1 - \frac{96}{\pi^4} \sum_{p=0}^{\infty} 
\exp \left[ - \frac{\pi^2 }{4} (2p+1)^2 \frac{t}{\tau} \right]
\frac{1}{(2p+1)^4}  \right)   \ .
\label{msd1d}
\end{eqnarray}
As expected, it saturates to  $\langle \delta x^2 (t) \rangle  =L^2/6 $ 
in the long-time limit $t \gg \tau$. More interesting is the behaviour of the MSD at short
times. Indeed, writing the Taylor expansion of the exponential
up to first order, we get
\begin{equation*}
\langle \delta x^2 (t) \rangle  =  \frac{L^2}{6}  - 
\frac{16 L^2}{\pi^4} \sum_{p=0}^{\infty}  \frac{ 1 }{(2p+1)^4}  +
\frac{16 Dt}{\pi^2} \sum_{p=0}^{\infty}  \frac{ 1 }{(2p+1)^2}
   + \Or \left(\frac{t^2}{\tau^2}\right)  \ .
\end{equation*}
From the well-known results of the series~\cite{arfkenbook} 
\begin{equation*}
\sum_{p=0}^{\infty}\frac{1}{ (2p+1)^{2}}= \frac{\pi^2}{8}  \qquad \mbox{and} \qquad
\sum_{p=0}^{\infty} \frac{1}{(2p+1)^{4}}= \frac{\pi^4}{96}
\end{equation*}
we find that the MSD does not depend on the size $L$ of the domain at short times, and
recover in this limit the celebrated result
\begin{equation}
\langle \delta x^2 (t) \rangle =  2Dt   \qquad t \ll \tau  \ .
\label{shortmsd1d}
\end{equation}

%
%

\section{Confined diffusion in a circular domain}
\label{seccircle}

\subsection{Green's function in 2D}

If the previous series  are easily evaluated, the corresponding
problem in two dimensions is more involved. 
In this section, we focus on the random motion of a tracer particle in a 2D circular domain
of radius  $a$.
In this geometry,  the Laplacian  reads
\begin{equation}
 \nabla^2    =\frac{1}{\rho}\frac{\partial }{\partial \rho} \rho \frac{\partial }{\partial \rho} +
\frac{1}{\rho^2} \frac{\partial^2 }{\partial \varphi^2}   
\label{laplace}
\end{equation}
where  $\mathbf{r}=(\rho, \varphi) $ are the coordinates of the particle.
The eigenfunctions of the radial part of $\nabla^2$ are the Bessel functions
of the first kind  $J_n$, which are solutions of the following differential equation~\cite{watsonbook}
\begin{equation}
x^2J''_n(x)+xJ'_n(x)+(x^2-n^2)J_n(x)=0   \ .
\label{edjn}
\end{equation}
The reflecting boundary conditions on the frontiers of the domain involve vanishing radial derivatives.
It can be shown that for $n \in \mathbb{N}$, $J'_n$ has an infinite number of real zeros, 
all of which are simple with the possible exception of $x=0$~\cite{watsonbook}.
In what follows, we shall note $\alpha_{nm}>0$ the $m^{th}$ positive root,  
$J'_n(\alpha_{nm})=0$, the zeros being 
arranged in ascending order of magnitude: $0< \alpha_{n1}<\alpha_{n2}<\ldots $. 
The Green's function of the problem is then given by the following expression~\cite{carslawbook}
\begin{eqnarray}
\fl G\left( \mathbf{r}, \mathbf{r}', t \right) = \frac{1}{\pi a^2}  + &  \frac{1}{\pi a^2} 
\sum_{n=-\infty}^{+\infty} \cos n(\varphi -\varphi')  \nonumber \\
& \times \sum_{m=1}^{\infty}   
\frac{   \alpha_{nm}^2   }{\alpha_{nm}^2-n^2 }
  \exp \left[- \alpha_{nm}^2 \frac{t}{\tau} \right]  
 \frac{J_n \Big( \displaystyle{ \alpha_{nm}\frac{\rho}{a}} \Big)  J_{n} \Big( \displaystyle{ \alpha_{nm}\frac{\rho}{a}} \Big)}
{ J_{n} \left(  \alpha_{nm} \right)^2}   
\label{green2d}
\end{eqnarray}
where we define the time-scale $\tau= a^2/D$.  
From this solution, we obtain the MSD
\begin{equation}
\langle \delta  \mathbf{r}^{2}(t) \rangle    =  a^2 \left( 1 -8 \sum_{m=1}^{\infty} \exp\left[ - \alpha_{1m}^{2} \frac{t}{\tau} \right]
\frac{ 1 }{\alpha_{1m}^2 \left(\alpha_{1m}^2-1\right)} \right)   \ .
\label{msd2d}
\end{equation}
with $\delta \mathbf{r}(t) = \mathbf{r}(t)- \mathbf{r} (0)$.
As expected, it
 saturates to $\langle \delta  \mathbf{r}^{2}(t) \rangle   =  a^2$ in the long-time limit
$t \gg \tau$. However, the short-time limit $t \ll \tau$ is more difficult to obtain. 
Writing the first terms of the Taylor expansion, we get
\begin{equation*}
\langle \delta  \mathbf{r}^{2}(t) \rangle   =  a^2  - 8 a^2 \sum_{m=1}^{\infty}
\frac{ 1 }{\alpha_{1m}^2 \left(\alpha_{1m}^2-1\right)}  +
8Dt  \sum_{m=1}^{\infty}
\frac{ 1}{   \left(\alpha_{1m}^2-1\right)}     + \Or \left(\frac{t^2}{\tau^2}\right)  .
\end{equation*}
Obviously, the short-time behaviour should  depend neither on the size nor on the particular
shape of the domain. Rather, one expects to find 
$\langle \delta  \mathbf{r}^{2}(t) \rangle    \sim   4Dt$ for $t \ll \tau$. 
We therefore have to get the numerical value of those sums in order to definitely check  this point,
and we now focus on generalized Rayleigh functions
in order to elucidate this question.

\subsection{Generalized Rayleigh functions}
\label{secsum}

At first sight, the evaluation of the infinite  series seems rather difficult to
achieve. But as we shall see here, an explicit expression 
for the zeros $\alpha_{1k}$ is actually not required.
We remind that the $\alpha_{1k}$'s are zeros of 
the Bessel function $J_1(x)$, the latter being given by~\cite{watsonbook,abrabook} 
\begin{equation}
J_1(x) = \sum_{n=0}^{\infty}      \frac{(-1)^n}{n! (n+1)!}    \left(\frac{x}{2}\right)^{2n+1}
= \frac{1}{2} \prod_{k=1}^{\infty} \left( 1-\frac{x^2}{\alpha^2_{1k}} \right)   
\label{expj1}
\end{equation}
The validity of the infinite product expansion follows from general properties of entire 
functions~\cite{arfkenbook}.
Let us then define  the generalized Rayleigh function~\footnote{Original 
Rayleigh functions are defined as 
the sum of reciprocal powers of zeros
of  Bessel functions. Here, we consider
zeros of {\it derivatives} of Bessel functions.} 
\begin{equation}
S_n  =   \sum_{k=1}^{\infty}
\left( \alpha_{1k} \right)^{-2n}    
\label{defsn}
\end{equation}
with $n \geq 1$ an integer.
Notice that the series is always well defined
since  $\alpha_{1k}\sim (k-\frac{1}{4}) \pi$
for large values of  $k$~\cite{abrabook}. 
To evaluate $S_n$, we follow a method originally developed by
Euler and Raleigh~\cite{watsonbook,muldoon1,muldoon2}. At a first step,
we introduce a new function $f(x)= J'_1(\sqrt{x})$, with $x \geq 0$.
The zeros of $f$ are $\xi_k=\alpha_{1k}^2$. From the representation of 
$J'_1$ as an infinite product~(\ref{expj1}), we write $f$ as
\begin{equation*}
f(x)=\frac{1}{2} \prod_{k=1}^{\infty} \left( 1-\frac{x}{\xi_k} \right)   \ .
\label{expf}
\end{equation*}
Next, we may differentiate this expression logarithmically to obtain
\begin{equation*}
\frac{f'(x)}{f(x)}  =   -  \sum_{k=1}^{\infty}  \frac{1/\xi_k}{ 1-x/\xi_k }  
 =   -  \sum_{k=1}^{\infty}  \frac{1}{\xi_k} \sum_{n=0}^{\infty}  \left( \frac{x}{\xi_k} \right)^n  \ .
\end{equation*}
The last series is absolutely converging for $0 \leq x  \leq 
1 $ (since $\alpha_{1k}>1$~\cite{abrabook}).
As a consequence, the order of summation can  be interchanged. 
Also evaluating the left-hand side of the  equality, we finally get
\begin{equation}
 \sum_{n=1}^{\infty} x^n   S_n  =  -\frac{\sqrt{x}  }{2}   . \frac{ J''_1\left(\sqrt{x} \right)}
{ J'_1\left(\sqrt{x} \right)}   \ .
\label{jsn}
\end{equation}

All the required quantities can be deduced from this result.
First, we  evaluate the series $\sum_{n=1}^{\infty} S_n$. This is done easily if we choose $x=1$
in~(\ref{jsn}), and together with~(\ref{edjn}) we find
\begin{equation}
\sum_{n=1}^{\infty}    S_n   =   \frac{1}{2}    \ .
\label{sn}
\end{equation}
Next, each series $S_n$ can be obtained  by identification of the coefficient of $x^n$. 
Given the Taylor expansion~(\ref{expj1}) of the Bessel function $J_1$, 
we readily get
\begin{equation}
 S_1   =   \frac{3}{8}   \ .
\label{s1}
\end{equation}
Though the  series   with $n \geq 2$  can 
be evaluated following the same procedure,
they will not be required for our purpose.

\subsection{Statistical properties}

Coming back to the original problem, we first need to calculate the series
\begin{equation*}
\sum_{m=1}^{\infty} \frac{ 1 }{ \left(\alpha_{1m}^2-1\right)}  =  \sum_{m=1}^{\infty} \frac{1}{\alpha_{1m}^2}  \
\frac{1}{1-\alpha_{1m}^{-2}}  
= \sum_{m=1}^{\infty} \sum_{n=1}^{\infty} \left( \alpha_{1m}\right)^{-2n}  \nonumber  \ .
\end{equation*}
Interchanging the order of summation, we obtain
\begin{equation}
 \sum_{m=1}^{\infty} \frac{ 1 }{ \left(\alpha_{1m}^2-1\right)}    =  \sum_{n=1}^{\infty} S_n =  \frac{1}{2}  \ .
\label{sum1}
\end{equation}
The other term is not more difficult to evaluate. Indeed, we have
\begin{equation*}
\sum_{m=1}^{\infty} \frac{ 1 }{\alpha_{1m}^2 \left(\alpha_{1m}^2-1\right)}  =  
 \sum_{m=1}^{\infty} \frac{ 1 }{ \left(\alpha_{1m}^2-1\right)}   - \sum_{m=1}^{\infty} \frac{1}{\alpha_{1m}^2} 
=  \sum_{n=1}^{\infty} S_n    -  S_1  \nonumber  
\end{equation*}
so that we find
\begin{equation}
\sum_{m=1}^{\infty} \frac{ 1 }{\alpha_{1m}^2 \left(\alpha_{1m}^2-1\right)} =  \frac{1}{8}  \ .
\label{sum2}
\end{equation}
We indeed recover the right behaviour
for the MSD at short times
\begin{eqnarray}
\langle \delta  \mathbf{r}^{2}(t) \rangle   &=& a^2  -8 a^2 \times \frac{1}{8} +
8Dt \times \frac{1}{2}  + \Or    \left( \frac{t^2}{\tau^2}  \right)  \nonumber \\
&=&   4Dt     \qquad  t \ll \tau  
\end{eqnarray}
as expected for $2D$ diffusion. 

At this point, it should be noticed that the method 
cannot be extended to systematically get each order of the expansion
of the MSD. Indeed, 
since the corresponding series do not converge,
it is not possible to interchange the order of summation
with the Taylor expansion.
Actually, this remark already holds for the problem in 1D --- see~(\ref{msd1d}),
and reflects the fact  the MSD is a {\it non-analytical}
function of time. Indeed,  it can be checked from the
large $s$ expansion of
the Laplace transform~(\ref{laplacemsd2d}) of the MSD that 
the next to leading term is proportional
to $t^{3/2}$. 

%
%

\section{Diffusion confined inside a sphere}
\label{secsphere}

\subsection{Green's function in 3D and statistical properties}

Similar questions arise for the diffusion of a particle confined in a sphere of radius a.
In this geometry, the Laplacian  reads
\begin{equation}
 \nabla^2    =\frac{1}{r^2}\frac{\partial }{\partial r} r^2 \frac{\partial }{\partial r} +
\frac{1}{r^2 \sin \theta } \frac{\partial }{\partial \theta} \sin \theta +
\frac{1}{r^2 \sin^2 \theta } \frac{\partial^2 }{\partial \phi^2}    \ .
\end{equation}
Eigenfunctions of the angular part of $\nabla^2$ are the spherical harmonics $Y_{lm}(\theta, \phi)$,
with eigenvalues $-l(l+1)$. Eigenfunctions of the radial part are the spherical Bessel functions
defined as $j_l(x)=(\pi /2)^{1/2} x^{-1/2} J_{l+1/2}(x)$, with $l$ a positive integer. They
 satisfy the following differential equation~\cite{watsonbook}
\begin{equation}
x^2j''_l(x)+2xj'_l(x)+(x^2-l(l-1))j_l(x)=0   \ .
\label{edsjn}
\end{equation}
We also need  to enforce reflecting boundary conditions on the surface of the sphere.
To this aim, we define $\beta_{ln}$ the (non-zero) zeros of the derivatives of the spherical Bessel
functions,  $j'_l(\beta_{ln})=0$.
They are arranged in ascending order of magnitude:  $0< \beta_{l1}<  \beta_{l2} < \ldots $. 
The normalization condition for the spherical Bessel functions is~\cite{arfkenbook}
\begin{equation}
\int_0^a  \rmd r r^2 j_l \left(\beta_{ln} \frac{r}{a} \right)^2 = 
\frac{a^3}{2 \beta_{ln}^2} \left( \beta_{ln}^2 -l(l+1) \right) j_l \left( \beta_{ln} \right)^2  
\label{norm}
\end{equation}
so that the Green's function for diffusion confined inside a sphere reads
\begin{eqnarray}
\fl G\left( \mathbf{r}, \mathbf{r}', t \right) = \frac{3}{4 \pi a^3} +   \frac{2}{a^3}
\sum_{l=0}^{+\infty} \sum_{m=-l}^l Y_{lm}^{ }(\theta,\phi)  Y_{lm}^{*}(\theta', \phi')  \nonumber \\
\times \sum_{n=1}^{\infty}   
\frac{   \beta_{ln}^2   }{\beta_{ln}^2-l(l+1) }  \exp \left[- \beta_{ln}^2 \frac{t}{\tau} \right]
 \frac{j_l \Big(\displaystyle{\beta_{ln} \frac{r}{a}} \Big)  j_{l} \Big( \displaystyle{ \beta_{ln} \frac{r'}{a}} \Big) }
{ j_{l} \left(  \beta_{ln} \right)^2}   \ .
\label{green3d}
\end{eqnarray}
Here, we used the same definition  $\tau=a^2/D$.

\subsection{Statistical properties}

From~(\ref{green3d}), we readily find $\langle r^2(t) \rangle = 3a^2/5$.
The MSD is then
\begin{equation*}
\left\langle \delta \mathbf{r}^2 (t) \right\rangle
 = \frac{6a^2}{5} - 2 \left\langle r(t) r(0) \cos \gamma \right\rangle  
\end{equation*}
with $\gamma$ the angle between the vectors $\mathbf{r}(t)$ and $\mathbf{r}(0)$. Rewriting $\cos \gamma
= \cos \theta \cos \theta' + \sin \theta \sin \theta' \cos (\phi - \phi')$ in terms of spherical
harmonics
\begin{equation}
\cos \gamma = \frac{4 \pi }{3} \sum_{m=-1}^1 Y_{1m}^{*}(\theta, \phi )  Y_{1m}^{ }(\theta', \phi' )  
\end{equation}
one can easily convince oneself that only the terms with $l=1$ give a non-zero contribution to
the mean value.  We get after some algebra 
\begin{equation}
\left\langle \delta \mathbf{r}^2 (t) \right\rangle  =  \frac{6a^2}{5} -12a^2 \sum_{n=1}^{\infty} 
\exp\left[ - \beta_{1n}^{2} \frac{t}{\tau} \right]
\frac{ 1 }{\beta_{1n}^2 \left(\beta_{1n}^2-2\right)}  \ .
\label{msd3d}
\end{equation}
As expected, the MSD saturates to $\langle \delta  \mathbf{r}^{2}(t) 
\rangle  \propto a^2$ in the long-time limit
$t \gg \tau$, whereas for short times $t \ll \tau$ we find
\begin{equation*}
\langle \delta  \mathbf{r}^{2}(t) \rangle  = 
\frac{6a^2}{5} -
12a^2 \sum_{n=1}^{\infty} \frac{ 1 }{\beta_{1n}^2 \left(\beta_{1n}^2-2\right)}
+12Dt \sum_{n=1}^{\infty} \frac{ 1 }{\beta_{1n}^2-2} 
    + \mathcal{O}\left(\frac{t^2}{\tau^2}\right)  \ .
\end{equation*}
The situation is similar to that in 2 dimensions, 
and we now  detail  the calculation of the infinite series.

\subsection{Short-time behaviour}

First, we rewrite the series that appear in the Taylor expansion of the MSD. 
After some elementary manipulations, we find
\begin{equation*}
\sum_{k=1}^{\infty} \frac{ 1 }{\beta_{1k}^2-2}  = 
\frac{1}{2}     \sum_{k=1}^{\infty}  \sum_{n=1}^{\infty} \left(\frac{\sqrt{2}}{ \beta_{1k}} \right)^{2n}  
= \frac{1}{2}   \sum_{n=1}^{\infty} \sigma_n  
\end{equation*}
where the series $\sigma_n$, with $n$ an integer $ \geq 1$, is defined as
\begin{equation}
\sigma_n = \sum_{k=1}^{\infty} \left(\frac{\sqrt{2}}{ \beta_{1k}} \right)^{2n}   \ .
\end{equation}
Note that $\beta_{11}>\sqrt{2}$~\cite{abrabook} so that all sums do converge. The other term  is given by
\begin{eqnarray*}
\sum_{k=1}^{\infty} \frac{ 1 }{\beta_{1k}^2 \left( \beta_{1k}^2-2\right)} &  = &
\frac{1}{2} \sum_{k=1}^{\infty} \frac{ 1 }{\beta_{1k}^2-2} -
\frac{1}{2} \sum_{k=1}^{\infty} \frac{ 1 }{\beta_{1k}^2}   \\
& = & \frac{1}{4} \left(   \sum_{n=1}^{\infty} \sigma_n - \sigma_1 \right)\ .
\end{eqnarray*}
The central point is to evaluate the series $\sigma_n$.
To this aim, it appears convinient to
consider the function  $j'_{1}(\sqrt{2x})$, whose zeros  are given by 
$\zeta_k=\beta_{1k}^2/2$. We then write $j'_{1}(\sqrt{2x})$ as an infinite product~\footnote{Compare
with  expression (11) of~\cite{muldoon1} with $a=0$,
$b= 2$, $c=-1 $, and $\nu=3/2$.}
\begin{equation*}
j'_{1}(\sqrt{2x})=\frac{1}{3} \prod_{k=1}^{\infty} \left( 1-\frac{x}{\zeta_k} \right)   
\end{equation*}
where the normalization factor comes from   $j'_1(0)=\frac{1}{3}$.
Taking the logarithmic derivative, we are directly lead  to
\begin{equation}
\sum_{n=1}^{\infty} x^n   \sigma_n  =  - \sqrt{\frac{x}{2}} . \frac{j''_1(\sqrt{2x})}{j'_1(\sqrt{2x})}   \ .
\label{seriesigma}
\end{equation}
From this result, we may draw the following conclusions:
\begin{itemize}
\item for $x=1$,  the numerical value of the right-hand side can be obtained from the differential equation~(\ref{edsjn})
satisfied by the spherical Bessel function. Since $2j''_1(\sqrt{2})+2\sqrt{2}j'_1(\sqrt{2})=0$,
it is found
\begin{equation}
 \sum_{n=1}^{\infty}\sigma_n = 1  \ .
\end{equation}
\item the series $\sigma_1$ is obtained by identification of the linear term in~(\ref{seriesigma}). 
The Taylor expansion of the modified Bessel function being $j_1(x)=\frac{x}{3}-\frac{x^3}{30} + \Or(x^5)$, 
we obtain
\begin{equation}
\sigma_1 = \frac{3}{5}  \ .
\end{equation}
\end{itemize}
Bringing everything together, we  obtain the expected result  for 3D diffusion, namely
\begin{eqnarray}
\langle \delta \mathbf{r}^2(t) \rangle &=& \frac{6a^2}{5}  -12a^2 \times \frac{1}{4}\left( 1-\frac{3}{5} \right) +
12 Dt \times \frac{1}{2}  + \Or \left( \frac{t^2}{\tau^2}  \right)  \nonumber \\
&=&   6Dt     \qquad  t \ll \tau  \ .
\end{eqnarray}

\section{Conclusion}
\label{concl}

To summarize, we have derived explicitly the
mean square displacement for confined diffusion in a disk or a sphere.
Since the results involve zeros of derivatives of Bessel functions, the numerical values of 
the relevant series are not easily obtained. To by-pass this difficulty, we used 
a powerful method  based on the expansion of entire functions as an infinite product.
This allowed us  to explicitly check that  $\langle \delta \mathbf{r}^2(t) \rangle \sim 2dDt$
in the short-time limit.  
Extension to more general domain shapes or space dimensions is straightforward, 
provided that the eigenfunctions of the Laplacian are known. 
This approach can also be adapted for adsorbing or mixed
boundary condition, and might as well be relevant  for the study of
polymer chains under  confinement. Indeed,
the partition function of a polymer satisfies a diffusion-like equation~\cite{edwardsbook},
where the time variable is replaced by 
the polymerization index $N$. 
In this case, the limit $t \ll \tau $ would correspond to 
 $R_g = \sqrt{Nb^2/6} \ll a$, with
$a$ the typical dimension of the cavity and 
$b$ the Kuhn length.

Finally, we  have to mention that the actual problem becomes more involved
in the (highly relevant) physical situation of  a confined colloidal suspension. 
Indeed, hydrodynamic interactions
with the bounding surface have to be accounted for 
if the particles are moving in a liquid,
and the diffusion coefficient is not expected to be constant
anymore. Instead, theoretical~\cite{brennerbook}
and experimental~\cite{lan,faucheux,dufresne} works have revealed  that
the mobility is in fact a function of position --- or, more precisely,
of the distance to the surface.
In addition, the diffusion coefficient may also be {\it time-dependent} when the 
bounding surface is {\it soft}, like a liquid-liquid interface or a fluid membrane. 
In some situations, it may even be  possible to observe anomalous diffusion, in which case 
the mean square displacement follows a power law
 $\langle \delta \mathbf{r}^2(t) \rangle \sim t^{\alpha}$,
 with $\alpha \neq 1$, over a notable range of
displacements and observation times~\cite{fradin,bickel}.

\ack

The author wish to thank L.~Cognet for usefull discussions. Correspondence
with E.~Oudet is also acknowledged.

\appendix
\section{Laplace transform of the MSD}

In this Appendix, we give the expression of the Laplace transform of
the MSD. Defining $\sigma_d(t)= \langle \delta {\bf r}^2(t) \rangle$, where
$d$ is the dimensionality of space, the Laplace transform
$\widetilde{\sigma}_d(s)= \int_0^{\infty} \rmd t \exp[-st] \sigma_d(t)$
is evaluated by solving directly the diffusion equation in terms
of the variable $s$ rather than $t$. Average values are then easily evaluated
since no infinite series are involved. Notice however that $\widetilde{\sigma}_d(s)$
can also be obtained by direct Laplace transform of the results~(\ref{msd1d}),
(\ref{msd2d}) and (\ref{msd3d}), though the method described in this article has
to be extended in order to evaluate the corresponding series. 
The result is
\begin{equation}
\widetilde{\sigma}_1(s) =
 \frac{2D}{s^2}  \left[  1-\frac{1}{ \sqrt{s \tau}}
\tanh \left(  \sqrt{s \tau}\right)  \right]  
\label{laplacemsd1d}
\end{equation}
for diffusion confined on a segment. For diffusion in a disk, one gets
\begin{equation}
\widetilde{\sigma}_2(s)
= \frac{4D}{s^2}\left[ 1 - \frac{1}{ \sqrt{s\tau}} \frac{I_1\left(\sqrt{s\tau}\right)}
{I'_1\left(\sqrt{s\tau}\right)}\right]  
\label{laplacemsd2d}
\end{equation}
with $I_1$ the modified Bessel function. Finally, the Laplace transform of the MSD
for diffusion confined inside a sphere is
\begin{equation}
\widetilde{\sigma}_3(s) = \frac{6D}{s^2}  
\left[ 1+ \frac{ \tanh \left(\sqrt{s\tau}\right) - \sqrt{s\tau}}{(s\tau +2) \tanh \left(\sqrt{s\tau}\right)
-2\sqrt{s\tau}} \right]  \ .
\label{laplacemsd3d}
\end{equation}
Asymptotic limits follow in a straightforward way.

\end{document}